\documentstyle[aps,prd,floats,epsf]{revtex}
\tighten
\draft
\begin{document}

\twocolumn[\hsize\textwidth\columnwidth\hsize\csname
@twocolumnfalse\endcsname
\title{Mass signature of supernova $\nu_\mu$ and $\nu_\tau$ neutrinos
in the Sudbury Neutrino Observatory}
\author{J.~F. Beacom\thanks{Electronic address:
        {\tt beacom@citnp.caltech.edu}} and
        P. Vogel\thanks{Electronic address:
        {\tt vogel@lamppost.caltech.edu}}}
\address{Department of Physics, California Institute of Technology\\
         Pasadena, CA 91125, USA}
\date{June 8, 1998}
\maketitle

\begin{abstract}

Core-collapse supernovae emit of order $10^{58}$ neutrinos and
antineutrinos of all flavors over several seconds, with average
energies of 10--25 MeV.  In the Sudbury Neutrino Observatory (SNO),
which begins operations this year, neutrinos and antineutrinos of all
flavors can be detected by reactions which break up the deuteron.  For
a future Galactic supernova at a distance of 10 kpc, several hundred
events will be observed in SNO.  The $\nu_\mu$ and $\nu_\tau$
neutrinos and antineutrinos are of particular interest, as a test of
the supernova mechanism.  In addition, it is possible to measure or
limit their masses by their delay (determined from neutral-current
events) relative to the $\bar{\nu}_e$ neutrinos (determined from
charged-current events).  Numerical results are presented for such a
future supernova as seen in SNO.  Under reasonable assumptions, and in
the presence of the expected counting statistics, a $\nu_\mu$ or
$\nu_\tau$ mass down to about 30 eV can be simply and robustly
determined.  If zero delay is measured, then the mass limit is {\it
independent} of the distance $D$.  At present, this seems to be the
best possibility for direct determination of a $\nu_\mu$ or $\nu_\tau$
mass within the cosmologically interesting range.  We also show how to
separately study the supernova and neutrino physics, and how changes
in the assumed supernova parameters would affect the mass sensitivity.

\end{abstract}


\vspace{0.5cm}]
\narrowtext


\section{Introduction}

As emphasized by Weinberg~\cite{Weinberg} and many others, whether or
not neutrinos have masses or other properties beyond the standard
model are questions which address some of the deepest issues in
particle physics.  Yet almost seventy years after they were proposed
by Pauli, most of the properties of neutrinos are defined only by
limits.  In particular their masses, if any, are unknown.  Results
from several experiments strongly suggest that neutrino flavor mixing
occurs in solar, atmospheric, and accelerator neutrinos, and proof of
mixing would be a proof of mass.  Direct searches for neutrino mass
yield only the following limits: $m_{\bar{\nu}_e} \lesssim 5$
eV\cite{Belesev}, $m_{\nu_\mu} < 170$ keV\cite{RPP}, and $m_{\nu_\tau}
< 24$ MeV\cite{RPP}.  With current techniques, it will be very
difficult to significantly improve these limits.  In fact, the
interesting mass scale is much lower than the latter two limits, and
is given by the requirement that neutrinos do not overclose the
universe (see \cite{Raffelt} and references therein):
\begin{equation}
\sum_{i=1}^3 m_{\nu_i} \lesssim 100 {\rm\ eV}\,.
\label{eq:cosmo}
\end{equation}
Neutrino masses exceeding this bound are allowed for unstable
neutrinos.

The most promising technique for direct determination of neutrino mass
below the cosmological bound seems to be from time-of-flight
measurements over astrophysical distances.  With the present
generation of detectors, neutrinos and antineutrinos of all flavors
from a Galactic supernova will be readily detectable.  Even a tiny
mass will make the velocity slightly less than for a massless
neutrino, and over the large distance to a supernova will cause a
measurable delay in the arrival time.  A neutrino with a mass $m$ (in
eV) and energy $E$ (in MeV) will experience an energy-dependent delay
(in s) relative to a massless neutrino in traveling over a distance D
(in 10 kpc) of
\begin{equation}
\Delta t(E) = 0.515
\left(\frac{m}{E}\right)^2 D\,,
\label{eq:delay}
\end{equation}
where only the lowest order in the small mass has been kept.  For a
$\nu_\mu$ or $\nu_\tau$ mass near the cosmological bound, the delay
for a single neutrino is of order seconds.  Because of the limit on
the $\bar{\nu}_e$ mass, this delay can be measured from the arrival of
the $\nu_\mu$ and $\nu_\tau$ events relative to the $\bar{\nu}_e$
events.  With the statistical power of many events, it is possible to
detect an average delay of order 0.1 s.  Since one expects a type-II
supernova about every 30 years in our Galaxy\cite{SNrate}, and since
supernova neutrino detectors are currently operating, there is a good
chance that this technique can be used to dramatically improve the
limits on the $\nu_\mu$ and $\nu_\tau$ masses.

In a previous paper~\cite{BV1}, we considered a future supernova at 10
kpc (approximately the distance to the Galactic center) as seen by the
SuperKamiokande (SK) detector.  Such a supernova will cause about 710
neutral-current excitations of $^{16}$O by $\nu_\mu$ and $\nu_\tau$
(and their antiparticles), followed by detectable gamma emission.  In
addition, about 8300 events are expected from $\bar{\nu}_e + p
\rightarrow e^+ + n$.  A $\nu_\mu$ or $\nu_\tau$ mass
would cause a delay of the average arrival time of the neutral-current
events as compared to the $\bar{\nu}_e$ events.
We have shown how to test the statistical significance of the
difference in average arrival times and how to extract the allowed
neutrino mass range.  Taking into account the finite statistics, we
concluded that with this signal at SK, one can reach a mass
sensitivity down to about 45 eV for the $\nu_\mu$ or $\nu_\tau$ mass.

In this paper, we consider the capabilities of the Sudbury Neutrino
Observatory (SNO).  For the same supernova, SNO will see in total
about 400 events caused by the neutral-current breakup of deuterons by
$\nu_\mu$ and $\nu_\tau$ (and their antiparticles).  As in
Ref.~\cite{BV1}, the technique used is to compare the average arrival
time of the (possibly massive) $\nu_\mu$ and $\nu_\tau$ events with
the average arrival time of the $\bar{\nu}_e$ events.  The sensitivity
of SNO for this measurement has been estimated
previously~\cite{SNO,Bahcall,Seckel,Krauss,Burrows}, with the claimed
minimal detectable mass ranging from 10 eV to 200 eV.  Here, we
present a detailed calculation of the mass sensitivity of SNO, taking
the finite statistics into account quantitatively.  While the
statistics are lower than for SK, the characteristic energy is lower
(and so the delay is larger), leading to a sensitivity to a $\nu_\mu$
or $\nu_\tau$ mass down to about 30 eV.

The problem of $\nu_\mu$ or $\nu_\tau$ mass determination with
supernova neutrinos has been discussed for other neutrino detection
reactions and using different analysis techniques.  Neutrino-electron
scattering in water-\v{C}erenkov detectors (e.g., SK) has been
considered in Refs.~\cite{Seckel,Krauss,Burrows,Minakata,Fiorentini}.
There have also been proposals to use neutral-current excitation of
various nuclei as a signal~\cite{Boron,SNBO,LAND,OMNIS}.  As shown
below, some of the previous estimates seem to be too optimistic in
that they assume a very sharp pulse of neutrinos in time, which makes
any delay more apparent.

In Section II, we describe the details of the supernova model, the
detector properties, and the neutrino detection signals.  In Section
III, we review our analysis technique and show our results for the
sensitivity of SNO to small $\nu_\mu$ or $\nu_\tau$ masses.  We also
consider how the mass sensitivity would be modified if the actual
supernova parameters differ from those assumed here.  In Section IV,
we discuss how the supernova parameters and neutrino properties can be
separately extracted from the same data.  In Section V, we summarize
our results.


\section{Production and detection of supernova neutrinos}


\subsection{Supernova neutrinos}

When the core of a large star ($M \ge 8 M_{\odot}$) runs out of
nuclear fuel, it collapses and forms a proto-neutron star with a
central density well above the normal nuclear density (for a review of
type-II supernova theory, see Ref.~\cite{Bethe}).  The total energy
released in the collapse, i.e., the gravitational binding energy of
the core ($E_B \sim G_N M_ {\odot}^2/R$ with $R \sim$ 10 km), is about
$3 \times 10^{53}$ ergs; about 99\% of that is carried away by
neutrinos and antineutrinos, the particles with the longest mean free
path.  The proto-neutron star is dense enough that neutrinos diffuse
outward over a timescale of several seconds, maintaining thermal
equilibrium with the matter.  When they are within about one mean free
path of the edge, they escape freely, with a thermal spectrum
characteristic of the surface of last scattering.  The luminosities of
the different neutrino flavors are approximately equal at all times;
see e.g., Ref~\cite{Woosley}.

Those flavors which interact the most strongly with the matter will
decouple at the largest radius and thus the lowest temperature.  As
explained in Ref.~\cite{BV1}, the $\nu_\mu$ and $\nu_\tau$ neutrinos
and their antiparticles, which we collectively call $\nu_x$ neutrinos,
have a temperature of about 8 MeV (or $\langle E \rangle \simeq$ 25
MeV).  The $\bar{\nu}_e$ neutrinos have a temperature of about 5 MeV
($\langle E \rangle \simeq$ 16 MeV), and the $\nu_e$ neutrinos have a
temperature of about 3.5 MeV ($\langle E \rangle \simeq$ 11 MeV); see
e.g., Refs.~\cite{Woosley,Janka}.  These are the temperatures used in
our analysis.  While there is some variation between models in the
actual values of the temperatures, all of the models have a
temperature hierarchy as above.  This is important for separating the
$\nu_x$ neutrinos from the $\nu_e$ and $\bar{\nu}_e$ neutrinos.  The
energy distributions are taken here to be Fermi-Dirac distributions,
characterized only by the temperatures given above.  More elaborate
models also introduce a chemical potential parameter to reduce the
high-energy tail of the Fermi-Dirac distribution; the effect of this
is considered below.

While some numerical supernova models have temperatures decreasing
with time, more recent models~\cite{Woosley} have temperatures
increasing with time.  This is a consequence of the electron fraction
and hence the opacities decreasing with time.  The real temperature
variation is probably not large (see e.g., Ref.~\cite{Woosley}).  A
well-motivated form for temperature variation may eventually be
obtained from the supernova $\bar{\nu}_e$ data or from more-developed
numerical models.  The analysis of this paper could be easily modified
to allow a varying temperature; until there is a compelling reason to
use a particular form, we simply use constant temperatures.

The neutrino luminosity rises quickly over a time of order 0.1 s, and
then falls over a time of order several seconds.  The luminosity used
here is composed of two pieces.  The first gives a very short rise
from zero to the full height over a time 0.09 s, using one side of a
Gaussian with $\sigma$ = 0.03 s.  The rise is so fast that the details
of its shape are irrelevant.  The second piece is an exponential decay
with time constant $\tau$ = 3 s.  The luminosity then has a width of
10 s or so, consistent with the SN 1987A observations.  The detailed
form of the neutrino luminosity is less important than the general
shape features and their characteristic durations.

This description of a supernova is consistent with theoretical
expectations, numerical supernova models, and the SN 1987A
observations.  With the next Galactic supernova, there will obviously
be great improvements in the understanding of the supernova neutrinos.
In Section IV, we discuss how to separately extract the supernova
parameters and neutrino properties from the same data.  Throughout the
paper, we assume that the distance to the supernova is $D = 10$ kpc.


\subsection{General form of the neutrino scattering rate}

Here we briefly summarize the notation of Ref.~\cite{BV1}.  Under the
assumption that the neutrino energy spectra are time-independent, the
double differential number distribution of neutrinos of a given flavor
(one of
$\nu_e,\bar{\nu}_e,\nu_\mu,\bar{\nu}_\mu,\nu_\tau,\bar{\nu}_\tau$) at
the source can be written as:
\begin{equation}
\frac{d^2 N_\nu}{dE dt_i} = f(E)\,\frac{L(t_i)}{\langle E \rangle}\,,
\label{eq:flux}
\end{equation}
where $E$ is the neutrino energy and $t_i$ is the emission time.  Here
$f(E)$ is the normalized thermal spectrum, $L(t_i)$ is the the
luminosity (energy flux per unit time), and $\langle E \rangle$ is the
(time-independent) average neutrino energy.  The double integral of
this quantity is the total number $N_\nu$ of emitted neutrinos of that
flavor.  This form is convenient since we assume that the luminosities
of the different flavors are approximately equal at every time $t_i$.
As stated above, the energy spectrum $f(E)$ is assumed to be a
Fermi-Dirac distribution, and the luminosity $L(t_i)$ is assumed to
have a very sharp rise and an exponential decline.

The arrival time of a neutrino of mass $m$ at the detector is $t = t_i
+ D + \Delta t(E)$, where $D$ is the distance to the source, and the
energy-dependent time delay is given by Eq.~(\ref{eq:delay}).  For
convenience, we drop the constant $D$.  Then the double differential
flux of neutrinos at the detector is given by
\begin{eqnarray}
\frac{1}{4\pi D^2} \frac{d^2 N_\nu}{dE dt} & = &
\frac{1}{4\pi D^2} \int dt_i\, \frac{d^2 N_\nu}{dE dt_i}
\delta(t - t_i - \Delta t(E)) \nonumber \\
& = &
\frac{1}{4\pi D^2} f(E)\,\frac{L(t - \Delta t(E))}{\langle E \rangle}\,.
\end{eqnarray}
Note that because of the mass effects, this is no longer the product
of a function of energy alone and a function of time alone.  The
scattering rate for a given neutrino reaction in SNO is
\begin{equation}
\frac{dN_{sc}}{dt} = N_{\rm{D}_2\rm{O}}\;
 n \int dE\,\sigma(E) \frac{1}{4\pi D^2} \frac{d^2 N_\nu}{dE dt}\,,
\end{equation}
where $N_{{\rm D}_2\rm{O}}$ is the number of heavy-water molecules in
the detector, $\sigma(E)$ the cross section for a neutrino of energy
$E$ on the target particle, and $n$ the number of targets per molecule
for the given reaction.  Using the results above, the scattering rate
(in s$^{-1}$) can be written:
\begin{equation}
\frac{dN_{sc}}{dt} = C
\int dE\,f(E) \left[\frac{\sigma(E)}{10^{-42} {\rm cm}^2}\right]
\left[\frac{L(t - \Delta t(E))}{E_B/6}\right]\,,
\label{eq:rate}
\end{equation}
\begin{equation}
C = 8.28
\left[\frac{E_B}{10^{53} {\rm\ ergs}}\right]
\left[\frac{1 {\rm\ MeV}}{T}\right]
\left[\frac{10 {\rm\ kpc}}{D}\right]^2
\left[\frac{{\rm det.\ mass}}{1 {\rm\ kton}}\right]
\,n\,.
\label{eq:C}
\end{equation}
In the above, $T$ is the spectrum temperature (where we assume
$\langle E \rangle = 3.15\,T$, as appropriate for a Fermi-Dirac
spectrum), and $f(E)$ is in MeV$^{-1}$.  For a light-water detector,
the initial coefficient in $C$ is 9.21 instead of 8.28.  For equal
luminosities in each flavor, the total binding energy released in a
given flavor is $E_B/6$.  Note that we ignore the prompt burst of
$\nu_e$ neutrinos, since these carry only of order 1\% of the total
energy.  When an integral over all arrival times is made, the
luminosity term in Eq.~(\ref{eq:rate}) integrates to one, giving for
the total number of scattering events:
\begin{equation}
N_{sc} = C
\int dE\,f(E) \left[\frac{\sigma(E)}{10^{-42} {\rm cm}^2}\right]\,.
\label{eq:total}
\end{equation}
For massless neutrinos, $\Delta t(E) = 0$, and in Eq.~(\ref{eq:rate})
the luminosity can be taken outside of the integral, so that the time
dependence of the scattering rate depends only on the time dependence
of the luminosity.


\subsection{The Sudbury Neutrino Observatory}

When it begins operations this year, the Sudbury Neutrino Observatory
(SNO) will be the first deuterium-based detector for neutrino
astrophysics.  The SNO detector is described in
Refs.~\cite{SNO,BGrate}.  Here we give a short summary of the
properties relevant to our analysis.  Deuterium is an excellent target
for neutrinos since both the charged- and neutral-current cross
sections are reasonably large.  The active part of the detector is 1
kton of pure D$_2$O, separated by an acrylic vessel from 1.7 kton of
light water.  This entire volume is viewed by $10^4$ phototubes which
can see about 1.4 kton of the light water with good
efficiency~\cite{McDonald}.  We assume that events in the D$_2$O can
be separated from events in the H$_2$O.

The proposed threshold for solar neutrino studies is 5 MeV (the
physics potential of the SNO solar neutrino studies is treated in
Ref.~\cite{Bahcall-Lisi} and references therein).  At this energy, the
contribution from the time-independent background is expected to be
small~\cite{BGrate}.  For a Galactic supernova, one expects several
hundred events over about 10 s, and a much higher background rate can
be tolerated, allowing a lower threshold.  For a threshold of 5 MeV,
the solar neutrino rate of about $10^{-4}$ s$^{-1}$ is a background
for supernova neutrinos.  Below 5 MeV, the background rate increases
very steeply.  From Ref.~\cite{BGrate}, we estimate that the threshold
for the supernova analysis can be lowered by a few MeV while keeping
the background contribution negligible.  This would ensure that almost
all low-energy gammas from neutron captures as well as electrons and
positrons from charged-current events are detected.

Electrons and positrons will be detected by their \v{C}erenkov
radiation, and gammas via secondary electrons and positrons.  It is
not possible for SNO to distinguish between electron, positrons, and
gammas of comparable energy.  The only way to detect the
neutral-current breakup of the deuteron is to detect the final
neutron.  There are three neutron detection techniques proposed for
SNO: $(n,\gamma)$ on deuterons in pure D$_2$O, giving a gamma of
energy 6.25 MeV; $(n,\gamma)$ on $^{35}$Cl in D$_2$O with MgCl$_2$
salt added, giving a gamma cascade of energy 8.6 MeV; and direct $n$
detection in $^3$He proportional counter tubes suspended into the
D$_2$O.  If either of the latter two techniques are used, there will
still be some contribution from neutron captures on deuterons.  In
this paper, as in all previous studies of the supernova capabilities
of SNO, we assume that the neutron detection efficiency is nearly
100\%.  In Section III, we show that a reduced neutron detection
efficiency has only a small effect on the mass sensitivity.  It
therefore makes very little difference which neutron detection
technique or techniques are in place when the supernova occurs.


\subsection{Neutrino signals in SNO}

Neutrinos and antineutrinos of all flavors cause the neutral-current
breakup of the deuteron:
\begin{equation}
\nu + d \rightarrow \nu + p + n\,,
\end{equation}
\begin{equation}
\bar{\nu} + d \rightarrow \bar{\nu} + p + n\,,
\end{equation}
with thresholds of 2.22 MeV.  There are about 35 $\nu_e$, 110
$\nu_\mu$, and 110 $\nu_\tau$ events from the first reaction, and 50
$\bar{\nu}_e$, 90 $\bar{\nu}_\mu$, and 90 $\bar{\nu}_\tau$ events from
the second reaction.  Electron neutrinos and antineutrinos in addition
cause the charged-current breakup of the deuteron:
\begin{equation}
\nu_e + d \rightarrow e^- + p + p\,,
\end{equation}
\begin{equation}
\bar{\nu}_e + d \rightarrow e^+ + n + n\,,
\end{equation}
with thresholds of 1.44 and 4.03 MeV, respectively.  There are about
80 events each from these two reactions.  The numbers of events are
summarized in Table~I.  The cross sections for all these reactions
have an energy dependence roughly of the form $\sigma(E) \sim (E -
E_{\rm th})^2$.  There have been many calculations of the
neutrino-deuteron cross sections~\cite{nud,Kubodera}, and they are now
rather well-determined.  In this paper, we use the tabulated cross
sections of Ref.~\cite{Kubodera}; the quoted uncertainty is 5\% or
less at the relevant energies.  All of the outgoing particles in the
deuteron breakup reactions are emitted approximately isotropically.

The neutral-current reactions are flavor-blind.  However, since
neutrinos and antineutrinos from a supernova have spectra with a
hierarchy of temperatures, the energy dependence of the rates favors
the higher-temperature flavors.  In particular, most (about 82\%) of
the events will be from $\nu_x$ neutrinos.  (Previous studies which
indicated a percentage of about 90\% used a higher $\nu_x$
temperature.)

\begin{table}[t]
\caption{Calculated numbers of events expected in SNO for a supernova
at 10 kpc.  The other parameters (e.g., neutrino spectrum temperatures)
are given in the text.  In rows with two reactions listed, the number of
events is the total for both.  The notation $\nu$ indicates the sum
of $\nu_e$, $\nu_\mu$, and $\nu_\tau$, though they do not contribute
equally to a given reaction, and $X$ indicates either $n + ^{15}$O or
$p + ^{15}$N.}
\begin{tabular}{l|r}
\multicolumn{2}{c}{Events in 1 kton D$_2$O}\\
\hline
$\nu + d \rightarrow \nu + p + n$ & 485 \\
$\bar{\nu} + d \rightarrow \bar{\nu} + p + n$ & \\
\hline
$\nu_e + d \rightarrow e^- + p + p$ & 160 \\
$\bar{\nu}_e + d \rightarrow e^+ + n + n$ & \\
\hline
$\nu + ^{16}{\rm O} \rightarrow \nu + \gamma + X$ & 20 \\
$\bar{\nu} + ^{16}{\rm O} \rightarrow \bar{\nu} + \gamma + X$ & \\
\hline
$\nu + ^{16}{\rm O} \rightarrow \nu + n + ^{15}{\rm O}$ & 15 \\
$\bar{\nu} + ^{16}{\rm O} \rightarrow \bar{\nu} + n + ^{15}{\rm O}$ & \\
\hline
$\nu + e^- \rightarrow \nu + e^-$ & 10 \\
$\bar{\nu} + e^- \rightarrow \bar{\nu} + e^-$ & \\
\hline\hline
\multicolumn{2}{c}{Events in 1.4 kton H$_2$O}\\
\hline
$\bar{\nu}_e + p \rightarrow e^+ + n$ & 365 \\
\hline
$\nu + ^{16}{\rm O} \rightarrow \nu + \gamma + X$ & 30 \\
$\bar{\nu} + ^{16}{\rm O} \rightarrow \bar{\nu} + \gamma + X$ & \\
\hline
$\nu + e^- \rightarrow \nu + e^-$ & 15 \\
$\bar{\nu} + e^- \rightarrow \bar{\nu} + e^-$ & \\
\end{tabular}
\end{table}

In order to estimate the delay, Eq.~(\ref{eq:delay}) can be evaluated
with a characteristic $\nu_x$ energy.  However, one should not use the
average energy $\langle E \rangle$, but rather a characteristic energy
that takes into account the weighting with the cross section as well.
For the neutral-current neutrino-deuteron reactions, this is shown
later to be about 32 MeV.  In addition, the fact that the neutrinos
have a spectrum of energies means that different values of $E$
contribute to the time delay, causing dispersion of the neutrino pulse
as it travels from the supernova.  It turns out that for the small
masses we are primarily interested in, these dispersive effects are
minimal.

Since the neutron capture time may be as large as several ms, and the
event rates may be high, concerns were raised in
Refs.~\cite{Krauss,Burrows} that events would overlap in time and that
it would therefore be difficult to distinguish charged-current and
neutral-current events.  Only during the first second or so are the
rates likely to be high enough that this can occur.  In the first
second, there are about 150 neutral-current events and only about 40
charged-current events.  Further, if the neutron mean free path is
less than the diameter of the acrylic vessel, then it will be possible
to use spatial information to distinguish events.  In any case, the
possible contamination of the neutral-current rate is very small, and
the mass sensitivity will not be significantly affected.

For the solar neutrino studies, the electron from the charged-current
$\nu_e$ reaction has a low energy and can be confused with a gamma
from a neutron-capture event.  Since the supernova $\nu_e$ energy is
much higher, the electron energy in a charged-current reaction is high
enough that only rarely can it be confused with a gamma.  Thus
neutrons can almost always be identified, either by direct detection
with a proportional counter, or by the energy of the subsequent gamma.
Positrons from charged-current events with $\bar{\nu}_e$ can be
identified by their high energy and coincidence with two neutrons.
The spectra of electron and positron energies will be broad, peaking
at about 15 and 20 MeV, respectively.

So far, we have discussed only the neutrino-deuteron reactions.  As
noted in Ref.~\cite{LVK}, the neutral-current excitation of $^{16}$O
into the continuum, followed by neutron or proton emission, can also
be detected.  About 30\% of the time, the A = 15 nucleus is left in an
excited state which decays by gamma emission, with gamma energies
between 5 and 10 MeV.  These gammas are detectable, as is the neutron.
The remaining 70\% of the time, the A = 15 nucleus is left in the
ground state.  In these cases, only the final states with a neutron
are detectable.  All of the outgoing particles are emitted
approximately isotropically.  Including the $^{16}$O events only has a
small effect on our final results.

There are also events from neutrino-electron scattering, which we
ignore.  These are forward-peaked, and we assume that they have been
removed with an angular cut.  A forward cone of half-angle 25 degrees
would contain almost all neutrino-electron scattering events, while
removing only 5\% of the isotropic events.


\section{Signature of a small neutrino mass}


\subsection{General description of the data}

For a massless neutrino ($\nu_e$ or $\bar{\nu}_e$) the time dependence
of the scattering rate is simply the time dependence of the luminosity
of that flavor.  For a possibly massive neutrino ($\nu_x$), the time
dependence of the scattering rate depends not only on the time
dependence of the luminosity of that flavor, but also on the delaying
effects of a mass.  Since the luminosities of the different flavors
are expected to be equal at all times, then if the $\nu_x$ is massive,
we can compare the $\nu_x$ and $\nu_e$ scattering rates to search for
a $\nu_x$ mass; see Eq.~(\ref{eq:rate}).  In Ref.~\cite{Burrows}, it
was proposed that the effects of a mass could be determined from the
time dependence of the neutral-current rate divided by the total rate,
which also has the effect of removing the the time dependence of the
luminosity from the scattering rate.  However, no quantitative
technique for extracting the allowed mass range was given.

In order to implement the comparison of scattering rates, we define
two rates: a Reference $R(t)$ containing only massless events, and a
Signal $S(t)$ containing some fraction of possibly massive events, and
some fraction of massless events (the possibly massive events cannot
be completely separated from the massless events).  We assume that
only $\nu_\tau$ is massive (this cannot be distinguished from the case
that only $\nu_\mu$ is massive).  The analysis can easily be repeated
for the case that both $\nu_\mu$ and $\nu_\tau$ are massive.  Because
of the quadratic dependence of the delay on the mass, this two-mass
case will look like a one-mass case with the larger mass unless the
two masses are nearly equal.

The Reference $R(t)$ can be formed in various ways.  The largest
sample of useful massless events will be the 7800 $\bar{\nu}_e$ events
from SK with $E_{e^+} > 10$ MeV.  Below 10 MeV, there are gammas from
the neutral-current excitation of $^{16}$O which cannot be separated.
We assume that SK and SNO will have synchronized clocks so that in
principle, such a sharing of data will be possible.  One can also use
the 340 $\bar{\nu}_e$ events with $E_{e^+} > 10$ MeV in the light
water at SNO.  Because of the smaller number of counts, the
statistical error is larger and hence the mass sensitivity is slightly
worse.  This could be slightly improved by including the 160
charged-current events in the heavy water (a small fraction of the
low-energy events would again have to be cut).

The primary component of the Signal $S(t)$ is the 485 neutral-current
events on deuterons.  With the temperatures assumed here, these events
are 18\% ($\nu_e + \bar{\nu}_e$), 41\% ($\nu_\mu + \bar{\nu}_\mu$),
and 41\% ($\nu_\tau + \bar{\nu}_\tau$).  The flavors of the
neutral-current events of course cannot be distinguished.  Therefore,
under our assumption that only $\nu_\tau$ is massive, there is already
some unavoidable dilution of the expected delaying effect of a mass.
The Signal $S(t)$ should also contain all events below about 10 MeV
which cannot otherwise be removed.  There are 35 neutral-current
events on $^{16}$O in the heavy water.  Because of the high threshold
for this reaction, these events are 50\% each of ($\nu_\mu +
\bar{\nu}_\mu$), and ($\nu_\tau + \bar{\nu}_\tau$).  Finally,
low-energy charged-current events must also be included.  Because of
the high energies of supernova neutrinos, only a small number (about
15) of electrons must be included in the Signal $S(t)$.  No positrons
need be included since those events can be identified by their two
accompanying neutrons.  The inclusion of the $^{16}$O and
charged-current events only slightly changes the shape of $S(t)$, and
barely changes the final mass limit.  The total number of events in
$S(t)$ is 535, of which still 41\% are possibly massive.

If the Signal $S(t)$ contains some fraction of massive events, the
shape of the scattering rate will be delayed and broadened.  Relative
to the Reference $R(t)$, there is a deficit of events at early times
and an excess at late times.  In the test for a mass, we test for this
characteristic distortion in the shape.  In Fig.~1, $S(t)$ is shown
under different assumptions about the $\nu_\tau$ mass.  The shape of
$R(t)$ is exactly that of $S(t)$ when $m_{\nu_\tau} = 0$, though the
number of events in $R(t)$ will be greater than in $S(t)$ if the SK
$R(t)$ is used and comparable if the SNO $R(t)$ is used.  For a very
large $\nu_\tau$ mass, the massless and massive components of $S(t)$
would completely separate in time.  In Fig.~1, note that for $m = 100$
eV, almost all of the massive events are delayed beyond 1 s.

\begin{figure}[t]
\epsfxsize=3.25in \epsfbox{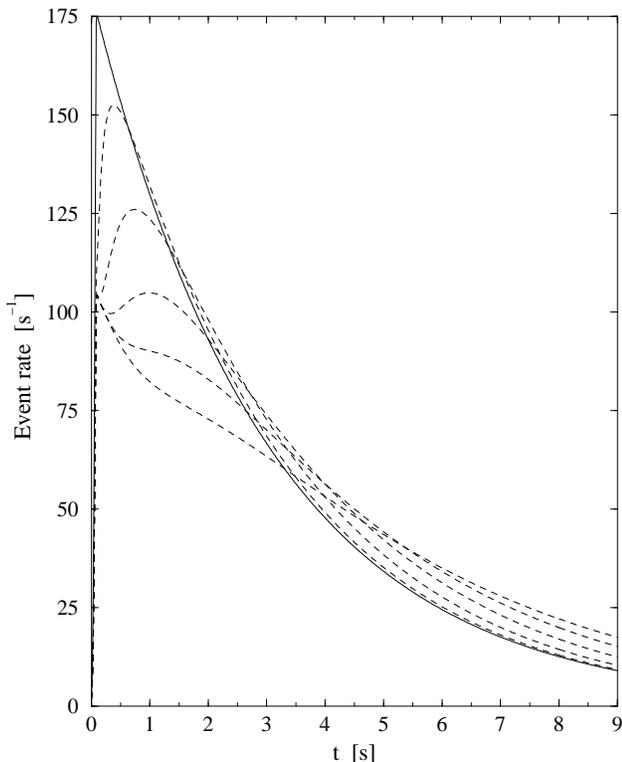}
\caption{The expected event rate for the Signal $S(t)$ at SNO in the
absence of fluctuations for different $\nu_\tau$ masses, as follows:
solid line, 0 eV; dashed lines, in order of decreasing height: 20, 40,
60, 80, 100 eV.  Of 535 total events, 100 are massless ($\nu_e +
\bar{\nu}_e$), 217.5 are massless ($\nu_\mu + \bar{\nu}_\mu$), and
217.5 are massive ($\nu_\tau + \bar{\nu}_\tau$).  These totals count
events at all times; in the figure, only those with $t \le 9$ s are
shown.}
\end{figure}

The rates $R(t)$ and $S(t)$ will be measured with finite statistics,
so it is possible for statistical fluctuations to obscure the effects
of a mass when there is one, or to fake the effects when there is not.
From Fig.~1, and Poisson statistics, one can easily get a rough idea
of how finely the mass can be determined from the difference between
$R(t)$ and $S(t)$.  Note that if the SK $R(t)$ is used, the
fluctuations in $R(t)$ when scaled down to the number of events in
$S(t)$ will be small, and that if the SNO $R(t)$ is used, the
fluctuations in $R(t)$ will be comparable to those in $S(t)$.

In this paper, we determine the mass sensitivity in the presence of
the statistical fluctuations by Monte Carlo modeling.  We use the
Monte Carlo to generate representative statistical instances of the
theoretical forms of $R(t)$ and $S(t)$, so that each run represents
one supernova as seen in SNO.  For $R(t)$, we pick a Poisson random
number from a distribution with mean given by the expected number of
events.  This determines the number of events for a particular
instance.  We then use an acceptance-rejection method to sample the
form of $R(t)$ until that number of events is obtained.  This gives a
statistical instance of $R(t)$, representative of what might be seen
in a single experiment.  A similar technique is used to generate an
instance of $S(t)$.  The massless and massive components of $S(t)$ are
sampled separately, and then added together.

In Ref.~\cite{BV1}, we considered two different tests of the shape
distortion of $S(t)$ relative to $R(t)$.  The first was a $\chi^2$
test.  A large value of $\chi^2$ between statistical instances of
$S(t)$ and $R(t)$ would indicate that they were likely to have been
drawn from different distributions, and this would be taken as
evidence of a $\nu_\tau$ mass.  However, this test is non-specific to
testing for a mass, i.e., it is sensitive to any difference between
$S(t)$ and $R(t)$.  In addition, the $\chi^2$ test requires binning in
time, which washes out the effects of small masses.  The second test,
and the one advocated there, was a test of the average arrival time
$\langle t \rangle$.  Any massive component in $S(t)$ will always
increase $\langle t \rangle$, up to statistical fluctuations.  Besides
being more directly related to the mass effect, the $\langle t
\rangle$ technique is sensitive to somewhat smaller masses than the
$\chi^2$ technique, since no binning is required.  In order that the
results be believable, it is necessary that different reasonable
statistical techniques yield consistent results.  For this paper, we
used both techniques and verified that they gave similar results.
However, we present only the results of the $\langle t \rangle$
analysis.


\subsection{$\langle t \rangle$ analysis}

Given the Reference $R(t)$, the average arrival time is defined as
\begin{equation}
\langle t \rangle_R = \frac{\sum_k t_k}{\sum_k 1} =
\frac{\int_0^{t_{max}} dt\,t R(t)}{\int_0^{t_{max}} dt\, R(t)} \,.
\end{equation}
The summation form is used for real or simulated data sets, where the
sum is over events (not time bins) in the Reference with $0 \le t \le
t_{max}$.  The integral form would be used if the theoretical form
for the rate were given.  The starting time is assumed to be
well-defined.  With some $10^4$ events expected in SK and SNO combined,
and a risetime of order 0.1 s, this should not be a problem; the
definition used here amounts to calling the starting point that time
at which the $\bar{\nu}_e$ rate is about 1\% of its peak rate.  The
choice of $t_{max}$ is made as follows.  The effect of the finite
number of counts in $R(t)$ is to give $\langle t \rangle_R$ a
statistical error:
\begin{equation}
\delta\left(\langle t \rangle_R\right) =
\frac{\sqrt{\langle t^2 \rangle_R - \langle t \rangle^2_R}}{\sqrt{N_R}}\,,
\label{eq:tRerror}
\end{equation}
where both the width $\sqrt{\langle t^2 \rangle_R - \langle t
\rangle^2_R}$ and the number of events $N_R$ depend on $t_{max}$.  By
choosing a moderate $t_{max}$, the width of $R(t)$ can be restricted.
Such a choice will make the error on $\langle t \rangle_R$ as small as
possible given that almost all of the events are to be included.  Both
the starting time and $t_{max}$ were of course held constant over
different Monte Carlo runs.  For a purely exponential luminosity, and
$t_{max} \rightarrow \infty$, $\langle t \rangle_R = \sqrt{\langle t^2
\rangle_R - \langle t \rangle^2_R} = \tau$.

Given the Signal $S(t)$, the average arrival time is defined similarly
as
\begin{equation}
\langle t \rangle_S = \frac{\sum_k t_k}{\sum_k 1} =
\frac{\int_0^{t_{max}} dt\,t S(t)}{\int_0^{t_{max}} dt\, S(t)} \,,
\end{equation}
where naturally the sums are now over events in the Signal.  The
widths of $R(t)$ and $S(t)$ are similar, each of order $\tau = 3$ s
(the mass increases the width of $S(t)$ only slightly for small
masses.)  If the SK $R(t)$ is used, then the statistical error on
$\langle t \rangle_S$ is few times larger than that on $\langle t
\rangle_R$ since there are several times fewer events.  If the SNO
$R(t)$ is used, then the statistical errors on $\langle t \rangle_S$
and $\langle t \rangle_R$ are comparable.  Note that the errors on
$\langle t \rangle_R$ and $\langle t \rangle_S$ are uncorrelated.

The signal of a mass is that the measured value of $\langle t
\rangle_S - \langle t \rangle_R$ is greater than zero with statistical
significance.  From the Monte Carlo studies, $t_{max} = 9$ s was found
to be a very reasonable choice for the luminosity decay time $\tau =
3$ s; about 95\% of the data are then included while the width is
somewhat reduced.  The time-independent background events are
negligible.  For $t_{max} = 9$ s, $\langle t \rangle_R = 2.57$ s and
$\sqrt{\langle t^2 \rangle_R - \langle t \rangle^2_R} = 2.12$ s.  Near
$t_{max} = 9$ s, the significance of a delay, i.e., $\langle t
\rangle_S - \langle t \rangle_R$ divided by its statistical error, is
nearly maximal for the small masses we are considering.  However, the
results are not strongly dependent on the particular value of
$t_{max}$ used as long as it is reasonable (i.e., the vast majority of
events are contained).  Note that any shift in the starting time will
cancel in the difference $\langle t \rangle_S - \langle t \rangle_R$.

Using the above procedure, we analyzed $10^4$ simulated supernova data
sets for a range of $\nu_\tau$ masses.  For each of them, $\langle t
\rangle_S - \langle t \rangle_R$ was calculated and its value
histogrammed.  These histograms are shown in the upper panel of Fig.~2
for a few representative masses.  (Note that the number of Monte Carlo
runs only affects how smoothly these histograms are filled out, and
not their width or placement.)  These distributions are characterized
by their central point and their width, using the 10\%, 50\% (equal to
the average), and 90\% confidence levels.  That is, for each mass we
determined the values of $\langle t \rangle_S - \langle t \rangle_R$
such that a given percentage of the Monte Carlo runs yielded a value
of $\langle t \rangle_S - \langle t \rangle_R$ less than that value.
With these three numbers, we can characterize the results of complete
runs with many masses much more compactly, as shown in the lower panel
of Fig.~2.  Since the $\langle t \rangle_S - \langle t \rangle_R$
distributions are Gaussians, other confidence levels can easily be
constructed.  For convenience, the axes in the lower panel are
inverted from how the plot was actually constructed.  That is, given
an experimentally determined value of $\langle t \rangle_S - \langle t
\rangle_R$, one can read off the range of masses that would have been
likely (at these confidence levels) to have given such a value of
$\langle t \rangle_S - \langle t \rangle_R$ in one experiment.  From
the lower panel of Fig.~2, we see that SNO has a sensitivity to a
$\nu_\tau$ mass down to about 30 eV (rounded from 27.5 eV) if the SK
$R(t)$ is used, and down to about 35 eV if the SNO $R(t)$ is used.

\begin{figure}[t]
\epsfxsize=3.25in \epsfbox{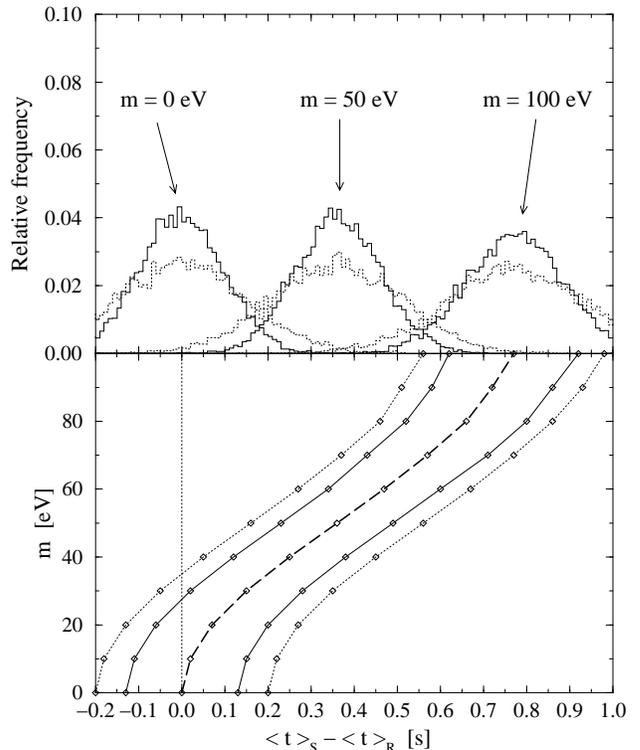}
\caption{The results of the $\langle t \rangle$ analysis for a massive
$\nu_\tau$, using the Signal $S(t)$ from SNO defined in the text.  In
the upper panel, the relative frequencies of various $\langle t
\rangle_S - \langle t \rangle_R$ values are shown for a few example
masses.  The solid line is for the results using the SK Reference
$R(t)$, and the dotted line for the results using the SNO $R(t)$.  In
the lower panel, the range of masses corresponding to a given $\langle
t \rangle_S - \langle t \rangle_R$ is shown.  The dashed line is the
50\% confidence level.  The upper and lower solid lines are the 10\%
and 90\% confidence levels, respectively, for the results with the SK
$R(t)$.  The dotted lines are the same for the results with the
SNO $R(t)$.  In this figure, $t_{max} = 9$ s and the time constant of
the exponential luminosity is $\tau = 3$ s.}
\end{figure}

We also investigated the dispersion of the event rate in time as a
measure of the mass.  A mass alone causes a delay, but a mass and an
energy spectrum also cause dispersion.  We defined the dispersion as
the change in the width $\sqrt{\langle t^2 \rangle_S - \langle t
\rangle^2_S} - \sqrt{\langle t^2 \rangle_R - \langle t \rangle^2_R}$,
where all integrals are as above defined up to $t_{max}$.  We found
that the dispersion was not statistically significant until the mass
was of order 80 eV or so; however, for such a large mass the
statistical significance of the change in $\langle t \rangle$ cannot
be missed.  For these large masses, dispersion does increase the width
of $S(t)$ and hence the error on $\langle t \rangle_S - \langle t
\rangle_R$; this is just becoming visible in Fig.~2.


\subsection{Sensitivity to the input parameters}

Since the parameters governing the supernova neutrino emission are not
perfectly known, it is worthwhile to examine the sensitivity of our
conclusions to their assumed values.  Most of this dependence can be
obtained analytically by extending the time integration limit to
$t_{max} \rightarrow \infty$.  Note that the mass sensitivity will be
slightly poorer than for $t_{max} = 9$ s used in the main analysis.

The characteristic delay is then
\begin{equation}
\langle t \rangle_S - \langle t \rangle_R \simeq
{\rm frac}(m > 0) \times 0.515 \left(\frac{m}{E_c}\right)^2 D\,,
\end{equation}
where ${\rm frac}(m > 0)$ is the fraction (about 41\%) of massive
events in $S(t)$ and the units are as in Eq.~(\ref{eq:delay}).  The
characteristic energy $E_c$ can be taken to be the peak of
$f(E)\sigma(E)$, or more accurately from
\begin{equation}
\langle\, \left(\frac{m}{E_c}\right)^2 \,\rangle =
\frac{\int dE\, f(E)\sigma(E) \left(\frac{m}{E}\right)^2}
{\int dE\, f(E)\sigma(E)}\,.
\end{equation}
The delay corresponding to $E_c$ is the delay of the centroid of
$S(t)$ relative to the centroid of $R(t)$, i.e., exactly $\langle t
\rangle_S - \langle t \rangle_R$.  For SNO, $E_c \simeq 32$ MeV.  The
statistical error on $\langle t \rangle_R$ is given by
Eq.~(\ref{eq:tRerror}), and the statistical error on $\langle t
\rangle_S$ is defined similarly.

Using the above, we can make a good estimate of the mass limit that
SNO would set if the delay were measured to be zero.  Ignoring the
error on $\langle t \rangle_R$, and taking $t_{max} \rightarrow
\infty$, we estimate the error to be $\tau/\sqrt{N_S} = 3/\sqrt{535} =
0.13$ s.  The 10\% and 90\% confidence levels as used in Fig.~2
correspond to $\pm 1.3 \times \delta\left(\langle t \rangle_R -
\langle t \rangle_S\right)$.  For our estimate, this has magnitude
0.17 s, close to the result in Fig.~2 for the SK Reference.  The mass
limit that will be placed if no delay is seen (using $D = 1$, i.e., 10
kpc) is
\begin{equation}
m_{limit} = E_c \,
\sqrt{\frac{1.3 \times {\rm error}}{ {\rm frac}(m > 0) \times 0.515 D}}
\simeq 30 {\rm\ eV}\,,
\end{equation}
in excellent agreement with the numerical result.  If the neutron
detection efficiency $\epsilon_n$ were not 100\%, then $N_S$ would be
reduced by $\epsilon_n$ and the error increased by
$1/\sqrt{\epsilon_n}$.  Therefore, the mass limit would be increased
only by $1/\sqrt[4]{\epsilon_n}$.

This formula can also be used to make an estimate of the mass limit
that can be obtained with the forward-peaked neutrino-electron
scattering at SK (which has the largest number of those events).
Since $\sigma(E) \sim E$, $E_c \simeq \langle E \rangle \simeq 25$
MeV.  The Signal $S(t)$ can be defined by the events in a forward cone
of half-angle 25 degrees, and the Reference $R(t)$ defined by the
events outside this cone.  Assuming that all of the neutrino-electron
scattering events are contained along with about 5\% of the isotropic
backgrounds, $N_S = 760$ and ${\rm frac}(m > 0) = 60/760 =
0.079$~\cite{BV1}.  The error is estimated to be $3/\sqrt{760} = 0.11$
s, and so $m_{limit} \simeq 50$ eV.  This is smaller than the recent
conclusion of Ref.~\cite{Fiorentini}, but the technique used here
includes a much greater portion of the data.

Continuing to suppose that the mass is very small, and that SNO will
simply place a limit, we can investigate the effects of varying the
input parameters.  We ignore dispersion and take the widths of $R(t)$
and $S(t)$ to be proportional to $\tau$.  If the cross section
$\sigma(E)$ depends on energy as $E^{\alpha}$ ($\alpha \sim 2$ for
$\nu + d$), then the characteristic energy $E_c \sim (2+\alpha)T$ and
the thermally-averaged cross section is proportional to $T^{\alpha}$,
where $T$ is the $\nu_x$ temperature.  We take ${\rm frac}(m > 0)$ to
be approximately constant and ignore the error on $\langle t
\rangle_R$; this is only valid for small deviations from $T = 8$ MeV.
With these assumptions we can determine how the key parameters affect
the result (for $\alpha = 2$).  The delay is
\begin{equation}
\langle t \rangle_S - \langle t \rangle_R \sim
\left(\frac{m}{T}\right)^2 D\,.
\end{equation}
The number of events is
\begin{equation}
N_S \sim \frac{1}{T} \frac{1}{D^2}\,T^2\,,
\end{equation}
and so the error is
\begin{equation}
\delta\left(\langle t \rangle_S - \langle t \rangle_R\right) \sim
\frac{\tau}{\sqrt{N_S}} \sim \frac{\tau D}{\sqrt{T}}\,.
\end{equation}
Therefore, the significance (number of sigmas) of the delay is
\begin{equation}
\frac{\langle t \rangle_S - \langle t \rangle_R}
{\delta\left(\langle t \rangle_S - \langle t \rangle_R\right)} \sim
\left(\frac{m}{T}\right)^2 \frac{\sqrt{T}}{\tau}\,.
\end{equation}
Remarkably, this is {\it independent} of $D$.  To place a mass limit
at a given confidence level, the number of sigmas is fixed.  Then the
mass limit that can be obtained if no delay is observed has the
following dependence:
\begin{equation}
m_{limit} \sim T^{3/4}\, \sqrt{\tau}\,, 
\label{eq:mlimit}
\end{equation}
also {\it independent} of $D$.

The $D$-independence can also be seen geometrically from Fig.~2.
Under $D \rightarrow 2D$, for example, the delay and the error each
become twice as large.  However, the point at which the upper
confidence level crosses the $m$-axis is unchanged.  If a nonzero delay
is measured, then the range of allowed masses does depend on $D$,
with larger distances corresponding to smaller masses.  The above
arguments hold for the relevant range 1--30 kpc; for smaller $D$ the
detector may be overwhelmed and for larger $D$ there are too few
events and the derivation is not valid.  Because of obscuration by
dust, it may be difficult to measure the distance to a future Galactic
supernova.  It is therefore rather important that this does not affect
the ability to place a limit on the $\nu_\tau$ mass.  That may not be
true for other analysis techniques; for example, the estimates of
Ref.~\cite{Seckel} do depend on $D$.  Note that a smaller distance
would  allow a better measurement of the temperatures and other
supernova parameters.

Now consider the effect of a change in the temperature $T_{\nu_x}$ on
the mass limit.  As indicated, the dependence is weak, with a higher
temperature making it slightly more difficult to limit the mass.  The
numerical results and the analytic estimate are shown in Fig.~3.  Even
under the large variation in $T_{\nu_x}$ of $\pm 2$ MeV, the mass
limit changes only by about $\pm 5$ eV.

\begin{figure}[t]
\epsfxsize=3.25in \epsfbox{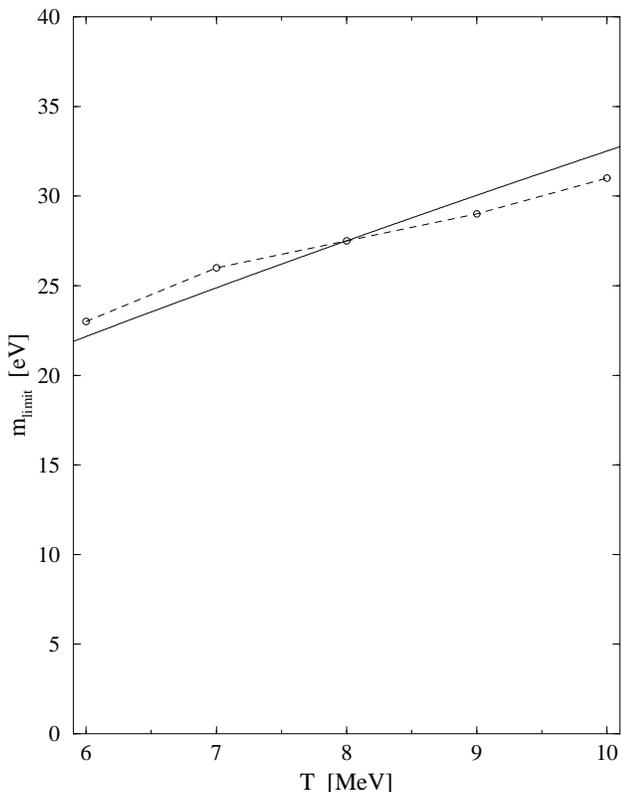}
\caption{The effect of a change in the $\nu_x$ temperature $T$ on the
mass limit that could be made using the SNO Signal and the SK
Reference.  The mass limit is defined as the mass at which the
appropriate upper confidence level intersects the $m$-axis in Fig.~2.
The solid line is the formula given in
Eq.~(\protect\ref{eq:mlimit}).  The full numerical results are
marked by circles, and connected with a dashed line to guide the eye.}
\end{figure}

The effect of changing the luminosity decay time constant $\tau$ is
even more straightforward and is obvious from Eq. (\ref{eq:mlimit}).
Thus, reducing that from $\tau = 3$ s to $\tau = 1$ s, as used in
Ref.~\cite{Krauss}, would improve the limit at SNO (using the SK Ref)
from about 30 eV to about 15 eV.  Using $\tau = 0.5$ s, as in
Ref.~\cite{LAND}, or $\tau = 0.3$ s, as in Ref.~\cite{OMNIS}, would
improve the limit to about 10 eV.  These values of $\tau$ were
estimated from appropriate figures in these references, and are only
approximations of the $L(t)$ timescale.  The numerical results and the
analytic estimate are shown in Fig.~4.  Using an unrealistically sharp
neutrino pulse makes the quoted mass limit very small.

\begin{figure}[t]
\epsfxsize=3.25in \epsfbox{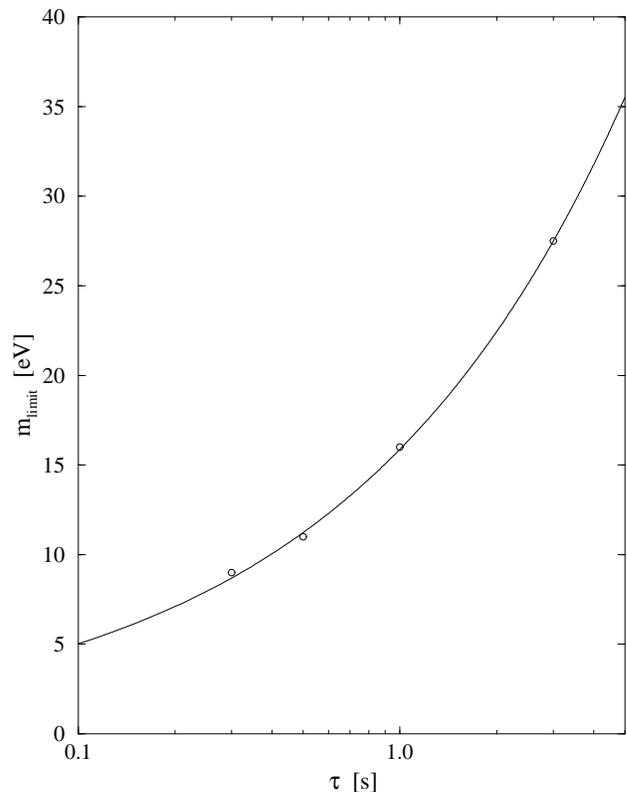}
\caption{The effect of a change in the luminosity decay constant
$\tau$ on the mass limit that could be made using the SNO Signal and
the SK Reference.  The mass limit is defined as the mass at which the
appropriate upper confidence level intersects the $m$-axis in Fig.~2.
The solid line is the formula given in Eq.~(\protect\ref{eq:mlimit}).
The full numerical results are marked by circles, at the points $\tau
= 3$ s (this paper), $\tau = 1$ s (Ref.~\protect\cite{Krauss}), $\tau =
0.5$ s (Ref.~\protect\cite{LAND}), and $\tau = 0.3$ s
(Ref.~\protect\cite{OMNIS}); see the text for explanation.}
\end{figure}

We also considered the effect of a chemical potential in the $\nu_x$
spectrum, which would reduce the high-energy tail.  We took $\mu =
3\,T$, and then chose $T = 6.31$ MeV to keep the average energy the
same as for the $\mu = 0$, $T = 8$ MeV case.  Because the
neutrino-deuteron cross section only depends quadratically on energy,
the effect is small.  About 10\% fewer events are obtained, but with a
delay about 10\% larger.  These lead to a change of order a few percent
in the mass sensitivity.  Because of their steeper energy
dependence~\cite{BV1}, the $\nu_x + ^{16}$O events would be more
affected by a chemical potential.


\section{Separate extraction of supernova and neutrino parameters}

The observation of the neutrino signal of a future Galactic supernova
will be extremely significant test of the physics involved.  It will
allow, among other things, determination of the imprecisely-known
supernova neutrino emission parameters.  In addition, we hope to be
able to use the same data to determine or constrain neutrino
properties.  In this section, we discuss how both of these goals can
be achieved simultaneously.  Throughout, we use standard values for
all parameters to numerically evaluate the expected precision.

The measured neutrino signal can be used to determine the temperatures
(and possibly the chemical potentials) of $\nu_e$, $\bar{\nu}_e$, and
$\nu_x$.  Ideally, we would like to know these parameters as a
function of time.  In addition, the data can be used to determine the
apparent source strength $E_B/D^2$ and the luminosity time profile.
Since neutrino oscillations could change the neutrino flavor between
emission and detection, their presence would make this task more
difficult.  Thus, we consider first the simpler situation without
neutrino oscillations.

\subsubsection{Without neutrino oscillations}

The $\nu_e$ and $\bar{\nu}_e$ neutrinos will be detected by the
charged-current reactions in which the energy of the outgoing electron
or positron will be measured.  Neglecting recoil corrections and
detector resolution, the relation $E_\nu = E_{e} + E_{th}$ allows one
to relate the measured electron or positron spectrum to the weighted
neutrino spectrum $f(E_\nu)\sigma(E_\nu)$ which appears in
Eq.~(\ref{eq:total}).  Since the charged-current cross sections are
well-known, the actual neutrino spectrum $f(E_\nu)$ can be obtained.
If this has the expected thermal shape, then the corresponding
temperatures $T_{\nu_e}$ and $T_{\bar{\nu}_e}$ (and possibly chemical
potentials) can be extracted.  The recoil corrections and detector
resolution can of course be taken into account by fitting the measured
electron or positron spectrum with an appropriately convolved trial
neutrino spectrum.  Note that these temperature determinations depend
only on the shape of the electron or positron spectrum.  The
normalization (the total number of events) can be used to determine
the apparent source strength $E_B/D^2$.

There are 7800 $\bar{\nu}_e + p \rightarrow e^+ + n$ events with
$E_{e^+} > 10$ MeV expected in SK, so that the average positron energy
can be determined to about 1\%.  Following the procedure above, one
should be able to determine the temperature $T_{\bar{\nu}_e}$ to a
comparable precision.  In this case one can also check for the
presence of a chemical potential, and for time variation of
$T_{\bar{\nu}_e}$.  The situation with $\nu_e$ is more difficult,
since there are only 80 events expected for the reaction $\nu_e + d
\rightarrow e^- + p + p$ in SNO.  (Note that $\nu_e$ has no
charged-current interaction with protons, that SNO is much smaller
than SK, and that $T_{\nu_e} < T_{\bar{\nu}_e}$.)  In this case, the
average electron energy can be determined to about 10\%; as above,
therefore a similar precision for the temperature $T_{\nu_e}$.  These
two measurements thus allow one to test $T_{\nu_e} < T_{\bar{\nu}_e}$,
as expected for the supernova temperature hierarchy.  In addition, the
two independent measurements of $E_B/D^2$ can be tested for
consistency with each other and with possible determinations unrelated
to neutrinos.  If $E_B/D^2$ is assumed to be known, then $T_{\nu_e}$
and $T_{\bar{\nu}_e}$ can also be determined from the total numbers of
charged-current events in SNO, as in Ref.~\cite{Balantekin}.

These measurements can also be compared with other available but
lower-statistics data.  For $T_{\bar{\nu}_e}$, other
hydrogen-containing detectors (MACRO, SNO, Borexino, and Kamland) will
have events from $\bar{\nu}_e + p \rightarrow e^+ + n$.  For
$T_{\nu_e}$, there are small numbers of events expected from
charged-current reactions on $^{12}$C in Borexino and Kamland, and
$^{16}$O in SK.  These consistency checks will be useful because the
different reactions may have different systematic errors.

The $\nu_x$ neutrino can be detected only by the neutral current
reactions.  Since the outgoing neutrino carries an unknown amount of
energy, it is not possible to determine the neutrino spectrum as
above.  Therefore, $T_{\nu_x}$ can only be extracted from the observed
total number of events, and must rely on the measured value of
$E_B/D^2$ and the assumption of luminosity equipartition.  For the
neutral-current reactions, the rate is independent of the neutrino
flavor.  If the supernova temperature hierarchy holds, then the
neutral-current events will be dominated by the $\nu_x$ neutrinos.
However, one cannot directly associate a flavor with the extracted
$T_{\nu_x}$.

The Signal at SK is expected to contain 710 events from $\nu_x +
^{16}$O.  Low-energy charged current events from $\bar{\nu}_e + p
\rightarrow e^+ + n$ can be confused with the neutral-current events;
these add 530 events to the Signal.  The latter events can be
subtracted from the total number of events observed in the Signal
since their expected number can be calculated using the parameters
measured from the higher-energy $\bar{\nu}_e$ data.  For the SK
Signal, since the dependence of the cross section on energy is
stronger, the 4\% precision on the number of events would translate to
a 1\% precision on $T_{\nu_x}$.  However, uncertainties on the cross
section itself should also be considered.  In principle, the
neutrino-electron scattering events at SK could also be used to
measure $T_{\nu_x}$; however, the temperature dependence of the number
of events is extremely weak.  The Signal at SNO is expected to contain
400 events from $\nu_x + d$, and 135 events from the sum of
neutral-current $\nu_e + \bar{\nu}_e$ reactions on deuterons,
neutral-current excitation of $^{16}$O, and low-energy $\nu_e$
charged-current reactions on deuterons.  Again, the numbers expected
for the latter can be calculated and subtracted from the total number
of events observed in the Signal.  The thermally-averaged cross
section for $\nu_x + d$ depends roughly on $T_{\nu_x}^2$, so by
Eq.~(\ref{eq:total}) the number of events is roughly proportional to
$T_{\nu_x}$.  Ignoring all other uncertainties, this would therefore
allow a determination of $T_{\nu_x}$ to about 5\%.  There may also be
$T_{\nu_x}$ measurements from $\nu_x + ^{12}$C at Borexino and
Kamland, particularly with the excitation of the 15.11 MeV state in
$^{12}$C.  If each neutral-current reaction were thoroughly
understood, it would be possible to use their different thresholds and
energy dependences to map out the $\nu_x$ spectrum.  At the very
least, it should be possible to extract a good value of $T_{\nu_x}$
and to make important consistency tests.

Therefore, all three temperatures can be extracted with reasonable
precision and the supernova temperature hierarchy $T_{\nu_e} <
T_{\bar{\nu}_e} < T_{\nu_x}$ experimentally verified.  As for the time
variation of the temperatures, it is likely that only the
$\bar{\nu}_e$ data will have sufficient statistics.  Those data, after
correcting for any temperature variation, will give the time profile
of the luminosity $L(t)$ and the characteristic timescale $\tau$.

Mass effects will not influence the temperature determination,
provided the mass is not too large, since a delay changes only the
time structure of the rate $S(t)$, not its normalization nor the
time-integrated energy spectra.  However, if $m_{\nu_x} \sim 1$ keV or
larger, some low-energy events will be lost in the time-independent
background due to their large delay, and the extracted temperatures
would be affected.  The considerations in this section, combined with
the result above that the mass limit is only weakly dependent on
$T_{\nu_x}$, shows that the temperatures and the $\nu_x$ mass can be
separately and robustly determined from the data.

\subsubsection{With neutrino oscillations}

Assuming that the presence of oscillations does not substantially
change the supernova dynamics, it is relatively straightforward to
separately search for oscillations and a mass delay.  The effects of
supernova neutrino oscillations (without a mass delay) were considered
for SNO in Ref.~\cite{Akhmedov}.  Since the numbers of $\nu_\mu$,
$\bar{\nu}_\mu$, $\nu_\tau$, and $\bar{\nu}_\tau$ are expected to be
the same, mixing among them have no effect on the number of $\nu_x$
events.  What propagates are the mass eigenstates; the flavor content
of the heavy mass eigenstate at the detector is irrelevant, and so
oscillations among the $\nu_x$ neutrinos do not affect the delay.

Now consider mixing between either $\nu_e$ and $\nu_\tau$ (or
$\nu_\mu$) or $\bar{\nu}_e$ and $\bar{\nu}_\tau$ (or $\bar{\nu}_\mu$).
We assume that the $\nu_\tau$ mass is large, which makes $\delta m^2$
large.  Large-angle, large-$\delta m^2$ mixing is ruled out by reactor
and accelerator experiments.  Small-angle, large-$\delta m^2$ mixing
is allowed; however, the effects are minimal unless there is an MSW
enhancement.  For a normal mass hierarchy, this can only happen for
mixing between $\nu_e$ and $\nu_\tau$, and at a very high density, of
order $10^{12}$ g/cm$^3$, which does occur in supernovae.  If there
are such oscillations, then some high-$T$ $\nu_\tau$ neutrinos will
become high-$T$ $\nu_e$ neutrinos.  As noted, the spectrum of
electrons from $\nu_e + d$ can be related to the spectrum of $\nu_e$
neutrinos.  For no oscillations, that would be a thermal spectrum with
$T_{\nu_e} = 3.5$ MeV; for complete oscillations, that would be a
thermal spectrum with $T_{\nu_e} = 8$ MeV (and a much greater number
of events).  For partial mixing, there would be two peaks.  The mixing
parameters and temperatures can thus be extracted from the measured
electron spectrum.  One should note that the charged-current reactions
$\nu_e + ^{12}$C at Borexino and Kamland and $\nu_e + ^{16}$O at SK
would have large numbers of events if such oscillations occur.  If
there is an inverted mass hierarchy, then mixing between $\bar{\nu}_e$
and $\bar{\nu}_x$ could have an MSW enhancement.  Then similar
considerations to those above could be used to examine the positron
spectrum from $\bar{\nu}_e$ reactions.  It would in fact be somewhat
easier, due to the larger numbers of events.  In either of these
cases, extracting the allowed $\nu_\tau$ mass range from the measured
value of $\langle t \rangle_S - \langle t \rangle_R$ requires some
care.  A given delay could be caused by a $\nu_\tau$ mass and no
mixing or by a larger mass and partial mixing.  Ideally, the mixing
parameters will be known from the considerations above, so that the
range of possible masses can be reasonably restricted.


\section{Conclusions and discussion}

One of the key points of our technique is that the abundant
$\bar{\nu}_e$ events can be used to calibrate the neutrino luminosity
of the supernova and to define a clock by which to measure the delay
of the $\nu_x$ neutrinos.  The internal calibration substantially
reduces the model dependence of our results, and allows us to be
sensitive to rather small masses.  Our calculations indicate that a
significant delay can be seen for $m = 30$ eV with the SNO data,
corresponding to a delay in the average arrival time of about 0.15 s.
Even though the duration of the pulse is expected to be of order 10 s,
such a small average delay can be seen because several hundred events
are expected.  Without such a clock, one cannot determine a mass limit
with the $\langle t \rangle_S - \langle t \rangle_R$ technique
advocated here, since the absolute delay is unknown.  Instead, one
would have to constrain the mass from the observed dispersion of the
events; only for a mass of $m = 150$ eV or greater would the pulse
become significantly broader than expected from theory.

Moreover, the technique used here allows accurate analytic estimates
of the results, so that it is easy to see how the conclusions would
change if different input parameters were used.  If the $\nu_\tau$
mass is very small, then the most probable measured delay is $\langle
t \rangle_S - \langle t \rangle_R = 0$.  In that case, one can only
place a limit $m_{limit}$ on the $\nu_\tau$ mass.  We have shown that
this limit is only weakly dependent on the $\nu_x$ temperature $T$ or
the presence of a chemical potential in the thermal spectrum, and is
{\it independent} of the distance $D$ for the reasonable range of
distances for a Galactic supernova.  The weak dependence of the
results on $T_{\nu_x}$ means that allowing a time variation of
$T_{\nu_x}$ would not significantly affect our conclusions.  The value
of $m_{limit}$ is sensitive to the timescale over which the luminosity
decreases.  If a very small value for the timescale is assumed, as is
sometimes the case in the literature, then one can obtain apparent
sensitivity to a very small $\nu_\tau$ mass.  However, such short
timescales are unreasonable, given the observed $\sim 10$ s duration
of the SN 1987A pulse.

The supernova parameters are not yet well-known.  However, the
sensitivity with which the neutrino properties can be determined using
the data from a future Galactic supernova depends upon the supernova
parameters assumed.  We have shown how the supernova and neutrino
physics can be separately studied using the same data, so that the
extracted neutrino parameters can be determined in an almost
model-independent way.

We assumed that only $\nu_\tau$ is massive (note that this cannot be
distinguished from the case that only $\nu_\mu$ is massive).  If no
delay is seen, and nothing further is known, then the mass limit
obtained would in fact apply to both the $\nu_\mu$ and $\nu_\tau$
masses.  If both are massive, then because of the quadratic dependence
of the delay on the mass, the one-mass case is recovered with the
larger mass unless the masses are similar.  If both are massive and
have the same mass, the delay would be increased by a factor 2, while
the error would be unchanged.  This would decrease the mass limit by a
factor $\sqrt{2}$ to about 20 eV.

If the $\nu_\tau$ mass is very large, then the pulse will be so
delayed and broadened that it will eventually disappear below the
time-independent background (which for this purpose includes solar
neutrinos).  Assuming that the background rate with a lowered
threshold is about $10^{-3}$ s$^{-1}$ (about 10 times the solar
neutrino rate), the maximal mass which can be seen in SNO is about $m
\sim 10$ keV, similar to SK~\cite{BV1}.  Above this mass, the neutrino
would be likely to decay over the distance to the supernova, as
pointed out in Ref.~\cite{Wolfenstein}.

Our previous paper~\cite{BV1} showed that using the neutral-current
excitation of $^{16}$O in SK gives a mass sensitivity down to about 45
eV.  In SK, one can also use the neutrino-electron scattering data,
and as noted above, that should have a sensitivity down to about 50
eV.  If no delay is seen, then SNO will place the best limit, about 30
eV, using neutral-current events on deuterons and calibrating the
$\bar{\nu}_e$ data from SK.  In that case, the two limits from SK
using neutral-current excitation of $^{16}$O and neutrino-electron
scattering can only confirm the result.  For those scattering rates, a
mass of 30 eV is insignificant, and may as well be zero, so that there
is nothing to be gained by combining those Signals with that from
neutrino-deuteron breakup at SNO.  On the other hand, if a significant
delay is seen, then there will be three independent determinations of
the allowed mass range.

In conclusion: We have presented a general method, including a
thorough statistical analysis, of extracting information about the
possible $\nu_\tau$ and $\nu_\mu$ masses from the future detection of
a Galactic supernova neutrino burst by the Sudbury Neutrino
Observatory.  When such an event in fact occurs, the existing mass
limits will be vastly improved and will approach, or cross over, the
cosmological bound.


\section*{ACKNOWLEDGMENTS}

This work was supported in part by the US Department of Energy under
Grant No. DE-FG03-88ER-40397.  J.F.B. was supported by a Sherman
Fairchild fellowship from Caltech.  We thank P.J. Doe, K.T. Lesko,
A.B. McDonald, and E.B. Norman of the SNO Collaboration for
discussions about the detector properties, and Y.-Z. Qian for
discussions about supernova neutrino oscillations.



\end{document}